\documentclass[letterpaper, 10 pt, conference]{ieeeconf}  % Comment this line out if you need a4paper

\IEEEoverridecommandlockouts                              % This command is only needed if 
\usepackage{amsmath,amssymb,amsfonts}
\usepackage{algorithmic}
\usepackage{graphicx}
\usepackage{textcomp}
\usepackage{multicol,lipsum}
\usepackage{mathtools}
\usepackage{cuted}
\usepackage{xcolor}

\usepackage[ruled,vlined]{algorithm2e}
\usepackage{algorithmic}
\usepackage{color,soul}
\usepackage[sorting=none, style=ieee,doi=false,isbn=false,url=false,eprint=false]{biblatex}
\usepackage{pgfplots}
\usepackage{tikz}
\usepackage{blindtext, rotating}
\usepackage{layouts}
\usepackage{printlen}
\usepackage{lipsum}
\usepackage{hyperref}
\usepackage{amsmath}
\usepackage[capitalise]{cleveref}
\usepackage{titlesec}
\usepackage{mathtools}
\usepackage[usestackEOL]{stackengine}
\usepackage{mleftright}

\AtEveryBibitem{\clearfield{month}}
\AtEveryBibitem{\clearfield{day}}

\usepackage{amsthm}

\newtheoremstyle{exampstyle}
{}
{}
  {} % Body font
  {} % Indent amount
  {\bfseries} % Theorem head font
  {.} % Punctuation after theorem head
  {.5em} % Space after theorem head
  {} % Theorem head spec (can be left empty, meaning `normal')

\theoremstyle{exampstyle} \newtheorem{definition}{Definition}[section]

\newcommand{\RomanNumeralCaps}[1]
{\MakeUppercase{\romannumeral #1}}

\title{\LARGE \bf
A Safe Control Architecture Based on a Model Predictive Control Supervisor for Autonomous Driving*
}

\author{Maryam Nezami$^{1}$, Georg Männel$^{1}$, Hossam Seddik Abbas$^{1}$ and Georg Schildbach$^{1}$% <-this % stops a space

\thanks{$^{1}$Institute for Electrical Engineering in Medicine,
        University of Lübeck, Lübeck, Germany
        {\tt\small \{maryam.nezami, ge.maennel, h.abbas, georg.schildbach\}@uni-luebeck.de} }
        \thanks{*H. S. Abbas is funded by the German Research Foundation (DFG), project number 419290163.}% 
}
\begin{document}
\maketitle
\thispagestyle{empty}
\pagestyle{empty}

%\titlespacing*{\section}{0pt}{0.2\baselineskip}{0.1\baselineskip}
%\titlespacing*{\section}{0pt}{0.2\baselineskip}{0.1\baselineskip}
%\titlespacing*{\subsection}{0pt}{0.3\baselineskip}{0.1\baselineskip}
%%%%%%%%%%%%%%%%%%%%%%%%%%%%%%%%%%%%%%%%%%%%%%%%%%%%%%%%%%%%%%%%%%%%%%%%%%%%%%%%
\begin{abstract}
This paper presents a novel, safe control architecture (SCA) for controlling an important class of systems: safety-critical systems. Ensuring the safety of control decisions has always been a challenge in automatic control. The proposed SCA aims to address this challenge by using a Model Predictive Controller (MPC) that acts as a supervisor for the operating controller, in the sense that the MPC constantly checks the safety of the control inputs generated by the operating controller and intervenes if the control input is predicted to lead to a hazardous situation in the foreseeable future invariably. Then an appropriate backup scheme can be activated, e.g., a degraded control mechanism, the transfer of the system to a safe state, or a warning signal issued to a human supervisor. For a proof of concept, the proposed SCA is applied to an autonomous driving scenario, where it is illustrated and compared in different obstacle avoidance scenarios. A major challenge of the SCA lies in the mismatch between the MPC prediction model and the real system, for which possible remedies are explored.

% This paper presents a new algorithm to assure safe obstacle avoidance for autonomous driving ground vehicles. The method is based on using a supervisor for the driver. The suggested supervisor is a model predictive controller (MPC). A safety detection event is proposed to avoid a dangerous maneuver. MPC takes over the vehicle instead of the driver when necessary. For the practical implementation of the MPC, we consider linear models of the vehicle. Therefore, we extend the approach to deal with the resulted mismatches between the vehicle's linear and nonlinear models. The effectiveness of the proposed methods is illustrated and compared in different obstacle avoidance scenarios. 
\end{abstract}
%%%%%%%%%%%%%%%%%%%%%%%%%%%%%%%%%%%%%%%%%%%%%%%%%%%%%%%%%%%%%%%%%%%%%%%%%%%%%%%%
\section{Introduction}
Safety-critical systems are a class of systems whose failure can cause damage to human health, life, property, or the environment. Examples of safety-critical systems include medical devices, aircraft flight control, and autonomous driving vehicles \cite{Knight2002}. Due to the increasing complexity in these applications, they make extended use of methods from Artificial Intelligence  (AI)  and Machine  Learning  (ML). However, the application of AI and ML based control algorithms is not suitable for many safety-critical systems, yet \cite{Fan2019}. Hence it is essential to consider approaches that can provide safety certificates in order to apply these methods to safety-critical systems.
% One approach AI researchers have paid more attention to recently is \textit{explainable} AI \cite{Ribeiro2016}, \cite{Park2018}, in order to avoid unwanted black-box behavior.  

% Therefore, these control algorithms to be applicable to a wider class of systems, their safe performance in a given range of physical and safety constraints must be checked \cite{Wabersich2018}.

% \sout{This is due to the inherent unreliability of these methods.} \textcolor{red}{(This would mean that it is generally impossible to make them reliable, and I wouldn't go that far! Maybe you want to mention here some research efforts / papers that try to make ML and AI safe, such as explainable AI.) 

The approach followed in this paper is to provide a safety mechanism for control systems, whose potentially unsafe control algorithm is treated as a black box algorithm. The proposed \textit{safe control architecture} (SCA) consists of a supervisor, which provides a safety certificate for the control input produced by an \textit{operating controller}. The operating controller refers to a general function block that contains potentially unsafe hardware or software elements, in particular, arbitrary AI and ML algorithms without safety guarantees or even a human operator acting as the controller. The safety certificate is generated by predicting the system state one step ahead if the control input were applied and checking whether this would lead to an unsafe situation. In cases where the control input is not certified, the supervisor activates a safety action. This action may be non-intrusive, such as a warning signal being displayed, or lead to a full take over by a redundant controller for the system. The proposed architecture is demonstrated in a case study for an autonomous driving vehicle but can be generalized to a wide class of safety-critical systems.

Safe decision making in safety-critical systems has gained considerable attention in recent years, e.g., \cite{Wabersich2018}, \cite{Akametalu2014}, \cite{Wabersich2017}. In \cite{Wabersich2018}, an optimization problem subject to the system constraints and a safe terminal set is solved to minimize the difference between an unsafe learning-based control input and an auxiliary control input. The safety of the learning-based control input is satisfied whenever the optimal cost is zero. In \cite{Akametalu2014}, safety based on reachability analysis has been employed to determine a safe operating region of the state space. Safety has been incorporated as a performance metric into a reinforcement learning algorithm to improve safety and learning. A safety framework has been introduced in \cite{Wabersich2017} to enhance learning-based and unsafe control strategies with safety certificates. It exploits the available data to deal with possible inaccuracies of using linear models. The main feature of the proposed architecture is that it is not restricted to a specific control problem, i.e., it can also be applied to tracking control problems, which are important in applications such as autonomous driving. 

Safe decision making is particularly important in the context of path planning and control in autonomous driving vehicles (ADV), and still, an open problem \cite{Baheri2019}. The safe decision making method in \cite{Gray2012a} is based on a predictive controller for the prevention of unintended road departure. Nevertheless, the method is difficult to be extended to other applications. A safe learning-based control framework has been introduced in \cite{Koller2019} by solving reinforcement learning tasks with state and input constraints. Safety of the system can be guaranteed while learning a given task. Yet, its applicability in real-time is questionable due to its computational complexity.

The contribution of this paper is threefold. Firstly, an SCA is proposed to guarantee the safe performance of safety-critical systems. Secondly, a Model Predictive Control (MPC) approach is introduced as the supervisor in the SCA. The Supervisor MPC is used to provide a safety certificate for the control input generated by any operating controller because, unlike non-optimization-based methods, it is able to provide a \textit{certificate of infeasibility}. Lastly, the application of the proposed SCA is demonstrated in a case study on autonomous driving for safe obstacle avoidance.

The paper consists of the following sections: Section \RomanNumeralCaps{2} describes the problem and the general safe control architecture using an MPC as the supervisor; Section \RomanNumeralCaps{3} describes the MPC setup for an obstacle avoidance scenario, and a method is presented to deal with model mismatch problems; Section \RomanNumeralCaps{4} illustrates the simulation setup; In Section \RomanNumeralCaps{5}, the results are demonstrated and discussed; Section \RomanNumeralCaps{6} presents the conclusion. 

% \sout{\textbf{Notations:} The set of of positive integers plus zero is denoted as $\mathbb{Z}_{0+}$.}

%%%%%%%%%%%%%%%%%%%%%%%%%%%%%%%%%%%%%%%%%%%%%%%%%%%%%%%%%%%%%%%%%%%%%%%%%%%%%%%%
\section{Supervisor MPC}
In this section, first, the model and some basic definitions concerning safety are stated. Next, the SCA is proposed. 
\subsection{Problem Description}
Consider the following nonlinear discrete-time system
\begin{equation}
	x_{k+1} = f(x_k,u_k),\quad x_0 = \bar{x}_0,
	\label{nonlinear_model}  
\end{equation} 
for some initial condition $\bar{x}_0 \in \mathbb{R}^n$. Here $f$ is a nonlinear function of $x_k, u_k$. The time step is denoted by $k \in \mathbb{Z}_{0+}$, where $\mathbb{Z}_{0+}$ is the set of positive integers. The state and input vectors of the system are $x_k \in \mathbb{R}^n$ and $u_k \in \mathbb{R}^m$, respectively. The system \eqref{nonlinear_model} is subject to the state and input constraints
\begin{equation}
	     x_k \in \mathcal{X}_k\, \forall k \in \mathbb{Z}_{0+}\;, \quad
        u_k \in \mathcal{U}_k\,\forall k \in \mathbb{Z}_{0+}\;.
\label{general_constraints}
\end{equation}
Here $\mathcal{X}_k$ is the state constraint set, and $\mathcal{U}_k$ is the input constraint set at time $k$.

% $\mathcal{X}_k$ is a polyhedron of the form $\mathcal{X}_k = \{ x \in \mathbb{R}^n | G_x x \leq h_x \}, G_x \in \mathbb{R}^{q_x \times n}, h_x \in \mathbb{R}^{q_x}$ and $\mathcal{U}_k$ is a polyhedron of the form $\mathcal{U}_k = \{ u \in \mathbb{R}^m | G^{\rm u} u \leq h^{\rm u}  \}, G^{\rm u} \in \mathbb{R}^{q_u \times m}, h^{\rm u} \in \mathbb{R}^{q_u}$.

In the example of this paper, the system is a passenger vehicle, which is a nonlinear system. The inputs of the system are the steering angle $\delta_k$ and the acceleration $a_k$. The input constraints are $  \delta_{\text{min}} \leq \delta_k \leq \delta_{\text{max}}$ and $ a_{\text{min}} \leq a_k \leq a_{\text{max}}$. To ensure smooth control inputs, constraints on the rate of change of $\delta_k$ and $a_k$ are enforced as $\Delta\delta_{\text{min}} \leq \delta_{k+1} -\delta_k \leq \Delta\delta_{\text{max}}$ and $ \Delta a_{\text{min}} \leq a_{k+1} - a_k \leq \Delta a_{\text{max}}$. The objective is to follow the road and to avoid any obstacles.
%\vspace*{-2mm}
\begin{definition}
		\normalfont{An \textbf{\emph{operating controller}} is an arbitrary control algorithm that generates the control inputs $u^{\rm o}_{k} \in \mathbb{R}^m$ to fulfill a control goal for system (\ref{nonlinear_model}).}
	\end{definition}
%\vspace*{-2mm}	
	
The basic goal is to assign a safety certificate for $u^{\rm o}_{k}$ at each time step $k$, or else prevent the system from using it. 

%\vspace*{-2mm}
\begin{definition} 
		\normalfont{At any time step $k$, system (\ref{nonlinear_model}) is said to be in a \textbf{\textit{safe state}} if there exists a feasible control sequence such that the constraints \eqref{general_constraints} are satisfied at all time steps in the future, i.e., $\exists \{ u_k , u_{k+1}, \cdots \} \quad \text{such that} \quad x_{k+j} \in \mathcal{X}_{k+j}, u_{k+j} \in \mathcal{U}_{k+j}, \forall j= 0,1,2, \cdots$. Conversely, the system (\ref{nonlinear_model}) is in an \textbf{\textit{unsafe state}} if the constraints \eqref{general_constraints} cannot be satisfied by the system dynamics by using any feasible control sequence, i.e., $\nexists \{ u_k , u_{k+1}, \cdots \} \quad \text{such that} \quad x_{k+j} \in \mathcal{X}_{k+j}, u_{k+j} \in \mathcal{U}_{k+j}, \forall j= 0,1,2, \cdots$. }
	\end{definition}
%\vspace*{-2mm}	

The safety certificate for $u^{\rm o}_{k}$ is attained if $u^{\rm o}_{k}$ leads the system \eqref{nonlinear_model} to a safe state in the next time step $k+1$. Then, $u^{\rm o}_{k}$ is said to be a \textit{safe control input}. Otherwise, $u^{\rm o}_{k}$ is an \textit{unsafe control input}. To avoid an unsafe state, the first step is to detect if the system is going from a safe state to an unsafe state. Therefore, the objective is to detect a \textit{safety event}.

% To avoid an unsafe state, just the detection of an unsafe state is not enough. The unsafe state has to be detected a little bit in advance, in order to allow the system to switch to a backup controller while it is still safe. 

%\vspace*{-2mm}
\begin{definition}
	\normalfont{A \textbf{\textit{safety event}} is when the application of $u^{\rm o}_{k}$ would drive the system \eqref{nonlinear_model} from a safe state to an unsafe state, e.g., due to an internal failure or functional error of the operating controller.}
\end{definition}
%\vspace*{-2mm}

A perfect detection of a safety event is a difficult task in practice. In reality, some \textit{reaction time} $t_r>0$ is required to intervene with $u^{\rm o}_{k}$ and to take over the control of the vehicle. Therefore, the safety event should be predicted. In this paper, for the purpose of a safety event prediction, a \textit{control supervisor} is proposed. Possible reactions to a safety event prediction depend on the safety concept. It may involve switching to a fully operational or degraded backup controller, transfer of the system to a pre-defined safe state, or a warning signal issued to a human supervisor.

%\vspace*{-2mm}
\begin{definition}
	\normalfont{ A \textbf{\textit{detection event}} is the prediction of a safety event by the control supervisor.} 
\end{definition}
%\vspace*{-2mm}

The detection event is based on all information that is available to the control supervisor, including $u^{\rm o}_{k}$, $x_k$, $\mathcal{X}_k$, $\mathcal{U}_k$, and the system dynamics. Due to imperfections, detection errors will be inevitable.

%\vspace*{-2mm}
\begin{definition}\label{detection_error} 
	\normalfont{ A \textbf{\textit{detection error type 1}} is when a safety event is predicted by the control supervisor, even though no safety event would entail without an intervention \textit{(false positive)}. A \textbf{\textit{detection error type 2}} is when a safety event occurs within the prediction time, even though no safety event is predicted by the control supervisor \textit{(false negative)}. }
\end{definition}
%\vspace*{-2mm}

%\vspace*{-2mm}
\begin{definition}
	\normalfont{ Assume that there is no detection error for a detection event. The \textbf{\textit{detection lead time}} $t_d$ is the time between the detection event and the safety event that would occur if $u^{\rm o}_{k}$ was not interrupted.  }
\end{definition}
%\vspace*{-2mm}

Ideally, $t_d = t_r$. If $t_d < t_r$, the violation of constraints cannot be avoided in the future. This means that a collision is bound to occur. If $t_d > t_r$, then this may lead to unnecessary or early intervententions with $u^{\rm o}_{k}$. This is usually considered as a performance loss.

\subsection{Safe control architecture}
\begin{figure} 
	\includegraphics{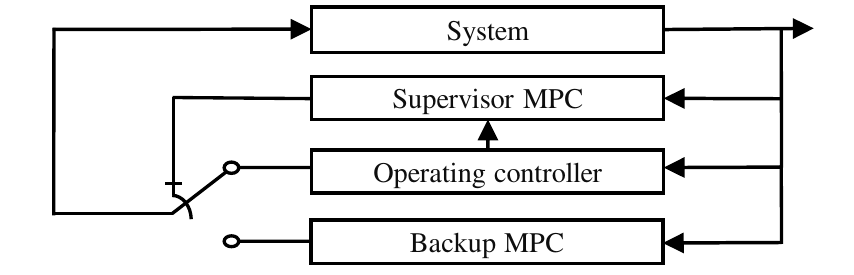}
	\caption{The block diagram of the proposed SCA by using an MPC as the supervisor and the backup controller}
	\label{block_diagram}
\end{figure}

In order to predict a safety event, a specific SCA is proposed that uses the MPC as the supervisor. In this architecture, a safety event is predicted by means of the MPC. Then, if a detection event occurs, another controller, in this case, called the ``Backup MPC'', takes over the system to avoid the unsafe state. The proposed SCA is illustrated in Figure \cref{block_diagram}.

% MPC is a control algorithm that uses a plant's model to predict the system's future evolution~\cite{Borrelli2006}. 
The proposed approach for the Supervisor MPC is to use the certificate of infeasibility of the underlying optimization problem to generate a detection event. Considering $x_{k}$ as the current system state, which is assumed to be measurable, an \textit{MPC-based detection event} is defined through the optimization problem  
\begin{subequations}\label{MPC}
	\begin{align} 
		\underset{U}{\text{min}}
		&\sum_{i=1}^{N} x^\top_{i|k} Q x_{i|k} + u^\top_{i-1|k} R u_{i-1|k}  \\
		 \text{s.t.} \;\;
		& x_{i|k} = f(x_{i-1|k},u_{i-1|k}), \quad \forall i = 1,\cdots,N,  \label{lin_model_in_MPC} \\
		& x_{0|k} = \hat{x}_{k+1}, \label{intial_condition_in_MPC} \\
		&  x_{i|k} \in \mathcal{X}_{i|k}, \quad \forall i = 1,2,\cdots,N,  \label{state_constraint_in_MPC}\\
		&  u_{i-1|k} \in \mathcal{U}, \quad \forall i = 1,2,\cdots,N, \label{input_contsraint_in_MPC}
	\end{align}
\end{subequations}
where \eqref{lin_model_in_MPC} is the nonlinear model from \eqref{nonlinear_model}. The initial condition is $x_{0|k} \in \mathbb{R}^n$. The predicted system state in the next step is $\hat{x}_{k+1}$, calculated as $\hat{x}_{k+1} = f(x_k,u^{\rm o}_{k})$. The tuning matrices are $Q \succeq 0 \in \mathbb{R}^{n\times n}$ and $R \succ 0 \in \mathbb{R}^{m\times m}$. The prediction horizon is $N$. The optimal control sequence found by solving the optimization problem \eqref{MPC} is $ U = \{ u^*_{0|k}, u^*_{1|k}, \cdots, u^*_{N-1|k} \} $.
% $\mathcal{X}_k$ is a polyhedron of the form $\mathcal{X}_k = \{ x \in \mathbb{R}^n | G_x x \leq h_x \}, G_x \in \mathbb{R}^{q_x \times n}, h_x \in \mathbb{R}^{q_x}$ and $\mathcal{U}_k$ is a polyhedron of the form $\mathcal{U}_k = \{ u \in \mathbb{R}^m | G^{\rm u} u \leq h^{\rm u}  \}, G^{\rm u} \in \mathbb{R}^{q_u \times m}, h^{\rm u} \in \mathbb{R}^{q_u}$.

%  The certificate is attained by the feasibility status of the MPC. The feasibility of MPC implies that there exists a control sequence such that all the constraints for all the future time are satisfied. Its infeasibility implies the opposite. Then, in feasibility of MPC is a safety event to be avoided. Since, as already stated in Section \RomanNumeralCaps{2}, a perfect detection of safety event is impossible, an \textit{MPC-based detection event} is proposed.

% In order to generate a detection event, the MPC is used as follows. Consider $x_{k}$ as the current system state, which is assumed to be measurable.. Accordingly, the operating controller generates a control input $u^{\rm o}_{k}$. Using the model \eqref{lin_model_in_MPC}, with the input $u^{\rm o}_{k}$, the system state in the next step can be predicted as $\hat{x}_{k+1}$. Then, the Supervisor MPC uses $\hat{x}_{k+1}$ as the initial condition $x_{0|k} = \hat{x}_{k+1}$ to generate a detection event based on the feasibility of MPC according to the following definition.
%\vspace*{-2mm}
\begin{definition} 
	\normalfont{An \textbf{\textit{MPC-based detection event} } is the fact that the optimization problem \eqref{MPC} is infeasible. In this case, the Supervisor MPC labels $u^{\rm o}_{k}$ as an unsafe control input and $u^{\rm o}_{k}$ should not be applied to the system; otherwise, if the problem is feasible, $u^{\rm o}_{k}$ is labeled as a safe control input and can be applied to the system.}
\end{definition}
%\vspace*{-2mm}

As long as the Supervisor MPC is not infeasible, the first element of the input sequence $\{ u^*_{0|k}, u^*_{1|k}, \cdots, u^*_{N-1|k}  \}$ is saved as a backup control input $u^{\text{backup}}_{k} = u^*_{0|k}$. At the moment $\tilde{k}$ when the Supervisor MPC is infeasible, the backup control input, which was calculated in the previous step, $u^{\text{backup}}_{\tilde{k}-1}$ is applied to the system. Thereafter, based on the proposed safety concept, the Backup MPC takes over the control of the system. In this case, the reaction time $t_r$ is one sampling time step. 

% In the following of this paper, it is assumed that the MPC controllers the system after a safety event prediction. Therefore, two working modes for MPC is proposed: \textit{"MPC in supervisory mode"} and \textit{"MPC in controlling mode"}. The first mode is to supervise the operating controller before a safety event prediction; the second mode is to control the system for avoiding the unsafe state after safety event prediction. 

%%%%%%%%%%%%%%%%%%%%%%%%%%%%%%%%%%%%%%%%%%%%%%%%%%%%%%%%%%%%%%%%%%%%%%%%%%%%%%%%
 \section{APPLICATION in AUTONOMOUS DRIVING}
The application of the proposed SCA in an autonomous driving scenario is demonstrated in this section. The scenario is for a vehicle to perform a safe obstacle avoidance maneuver (SOAM). At first, the MPC design is presented; next, problems arising for a SOAM due to the mismatch between the model and the vehicle dynamics are described, and eventually, a solution to fulfill a successful SOAM in the presence of model mismatches is presented.

\subsection{MPC design for a safe obstacle avoidance maneuver}
The design of the MPC problem \eqref{MPC} for a SOAM by a vehicle is presented in this subsection. To this end, a dynamic model in which the state variables are in terms of position and orientation error with respect to the road is utilized \cite{Rajamani2012}:
\begin{multline}
	\!\!\!\!\!\!\!\!\frac{\text{d}}{\text{d}t}\!\!\begin{bmatrix}
		\!e_y\! \\ \!\dot{e}_y\! \\ \!e_\psi\! \\ \!\dot{e}_\psi\! \\ \!e_x\! \\ \!\dot{e}_x\!
	\end{bmatrix}\!\!\!= 
	 \underbrace{\!\!\! \begin{bmatrix}
		\!	0   \!\!   &  \!\!    1   \!\!  &  \!\!   0   \!\!  &  \!\!   0 \!\!    &   \!\! 0  \!\!   &  \!\!   0 \!    \\
		\!	0   \!\!   &  \!\!  a  \!\!   & \!\!   b  \!\!    & \!\!   c  \!\!    &  \!\!   0   \!\!  & \!\!    0 \!    \\
		\!	0   \!\!   &  \!\!   0   \!\!  &  \!\!   0  \!\!   &  \!\!   1  \!\!   &   \!\!  0   \!\!  &  \!\!   0  \!   \\
		\!	0   \!\!   &  \!\!   d  \!\!   &  \!\!  e   \!\!   &   \!\! f    \!\!  &   \!\!  0   \!\!  &  \!\!   0  \!  \\
		\!	0   \!\!   & \!\!    0   \!\!  & \!\!    0   \!\!  & \!\!    0  \!\!   &   \!\!  0   \!\!  &   \!\!  1   \!  \\
		\!	0    \!\!  &   \!\!  0 \!\!    &  \!\!   0  \!\!   &   \!\!  0  \!\!   &   \!\!  0   \!\!  & \! \!   0    \! \\
	\end{bmatrix}\!\!\!}_{\triangleq A_1}  \overbrace{\!\!\!\!\!\!\!\!\!\!\!\!\!\!\begin{bmatrix}
	\!	e_y \!\\ \!\dot{e}_y\! \\\! e_\psi \!\\ \!\dot{e}_\psi \!\\ \!e_x \!\\\! \dot{e}_x\!
	\end{bmatrix}\!\!\!\!\!\!\!\!\!\!\!\!\!\!}^{ \triangleq x_1(t)}  
	\!\!\!+ \underbrace{\!\!\begin{bmatrix}
		\!	0 \!\!&\! 0 \! \\ \! \frac{2C_{\alpha f}}{m} \!\!&\!\! 0 \! \\
		\!	0 \!\!&\!\! 0 \! \\
		\!	\frac{2 C_{\alpha f} l_f}{I_z} \!\!&\!\! 0 \! \\
		\!	0 \!\!&\!\! 0 \! \\
		\!	0 \!\!&\!\! 1 \!
	\end{bmatrix}\!\!\!}_{\triangleq B_1} \overbrace{\!\!\!\begin{bmatrix}
	\!	\delta(t) \!\\ \!a(t)\!	\end{bmatrix}\!\!\!}^{\triangleq u_1(t)} \!\!+ \underbrace{\!\!\begin{bmatrix}
		\!	0 \!\!&\!\! 0 \!\\ \!g \!\!&\!\! 0\!  \\ \!0 \!\!&\!\! 0\! \\ \!h \!\!&\!\! 0\! \\ \!0 \!\!&\!\! 0 \!\\ \!0 \!\!&\!\! -1\!
	\end{bmatrix}\!\!\!}_{\triangleq E_1} \begin{bmatrix}
		\!\dot{\psi}_{\text{des}}(t)\! \\ \!a_{\text{ref}(t)}\!
	\end{bmatrix}\!\!,\!\!\!\!\!\!
	\label{eq_sixteen}
\end{multline}
where 
% $a,b,c,d,e,f,g$ and $h$ can be found in \cite{Rajamani2012} page $36$. 
\begin{equation*}
	\begin{split}
		a &= -\frac{2C_{\alpha f} + 2 C_{\alpha r}}{m V_x}, \quad b = -\frac{2C_{\alpha f} + 2 C_{\alpha r}}{m}, \\
		c &= -\frac{2 C_{\alpha f} l_f + 2 C_{\alpha r} l_r}{m V_x}, \quad d = -\frac{2 C_{\alpha f} l_f - C_{\alpha r} l_r}{I_z V_x}, \\
		e &=  \frac{2 C_{\alpha f} l_f - 2 C_{\alpha r} l_r}{I_z}, \quad f =  -\frac{C_{\alpha f}l^2_f + 2 C_{\alpha r} l^2_r}{I_z V_x}, \\
		g &= -\frac{2C_{\alpha f}-2C_{\alpha r} l_r}{m V_x} - V_x, \quad h = -\frac{2 C_{\alpha f} l^2_f + 2 C_{\alpha r} l^2_r}{I_z V_x}.
	\end{split}
\end{equation*}
% \begin{equation*}
%     a = -\frac{2C_{\alpha f} + 2 C_{\alpha r}}{m V_x}, \quad b = -\frac{2C_{\alpha f} + 2 C_{\alpha r}}{m},
% \end{equation*}
% \begin{equation*}
%     c = -\frac{2 C_{\alpha f} l_f + 2 C_{\alpha r} l_r}{m V_x}, \quad d = -\frac{2 C_{\alpha f} l_f - C_{\alpha r} l_r}{I_z V_x},
% \end{equation*}
% \begin{equation*}
%     e =  \frac{2 C_{\alpha f} l_f - 2 C_{\alpha r} l_r}{I_z}, \quad f =  -\frac{C_{\alpha f}l^2_f + 2 C_{\alpha r} l^2_r}{I_z V_x},
% \end{equation*}
% \begin{equation*}
%     g = -\frac{2C_{\alpha f}-2C_{\alpha r} l_r}{m V_x} - V_x, \quad h = -\frac{2 C_{\alpha f} l^2_f + 2 C_{\alpha r} l^2_r}{I_z V_x}.
% \end{equation*}
Here $e_y$, $e_\psi$, and $e_x$ represent the distance of the center of gravity (CoG) of the vehicle from the center-line of the road, the orientation error of the vehicle with respect to the road, and the longitudinal position error with respect to the center-line of the road, respectively. The offline generated control inputs are $\dot{\psi}_{\text{des}}$ and $a_{\text{ref}}$, calculated based on the derivative of the yaw angle of the desired road and the reference speed, respectively. The list of vehicle parameters and their units is shown in \cref{para-table}.\\
\begin{table}
	\caption{Vehicle parameters  used in vehicle dynamic modeling~\cite{Gottmann2018}}
	\label{para-table}
	\begin{center}
		\begin{tabular}{ c l c}
			\hline
			\textbf{Symbol} & \textbf{Parameter}& \textbf{Value}  \\
			\hline
			$C_{\alpha f}$ & Cornering Stiffness Front &  $153$\,kN$/$rad \\
			
			$C_{\alpha r}$ & Cornering Stiffness Rear & $191$ \,kN$/$rad \\
			
			$l_f$ & Distance CoG to Front Axle & $ 1.3$ \,m \\
			
			$l_r$ & Distance CoG to Rear Axle & $1.7$ \,m \\
			
			$I_z$ & Vehicle Yaw Inertia & $5250$ \,kgm$^2$ \\
			
			$V_x$ & Longitudinal Velocity of the Vehicle & $10$\,m$/$s \\
			
			$m$ & Vehicle Mass & $2500$\,kg \\
			\hline
		\end{tabular}
	\end{center}
\end{table}
In order to implement the constraints on the rate of change of $\delta$ and $a$, the linear model \eqref{eq_sixteen} is augmented with two integrators
\begin{multline} \label{new_rep_lin_model}
	\!\!\!\!\!\!\!\!
	\frac{\text{d}}{\text{d}t}\!\! \begin{bmatrix}
		x_1(t)  \\ u_1(t)
	\end{bmatrix} \! \! 
	=
	\underbrace{\!\!\mleft[
\begin{array}{c|c}
\! \! \!  A_1  \! & \! B_1 \! \! \!\\
  \hline
\! \! \!  \boldsymbol{0} \! & \! \boldsymbol{0} \!\! \! 
\end{array}
\mright] \!\!}_{\triangleq A_{\text{c}}} \;
    \underbrace{\!\!\begin{bmatrix}
	\!	x_1(t) \! \\ \! u_1(t) \!
	\end{bmatrix}\!\!}_{\triangleq x(t)} 
	 + 
	 \underbrace{\!\!\mleft[
\begin{array}{c}
 \!\!\! \boldsymbol{0} \!\!\! \\
  \hline
  \begin{matrix}\!\!\! 1 \!\!\ & \!\! 0 \!\!\!\\ \! \!\!\!0 \!\! & \!\! 1 \!\!\!  \end{matrix}
\end{array}
\mright]\!\!}_{\triangleq B_{\text{c}}} \;
	\underbrace{\!\!\!\begin{bmatrix}
		\!	\dot{\delta}(t)  \! \\ \! \dot{a}(t) \!
	\end{bmatrix}\!\!\!}_{\triangleq u(t)} 
	+
	\underbrace{\!\!\! \mleft[
\begin{array}{c}
 \!\!\! E_1 \!\!\! \\
  \hline
\!\!\! \boldsymbol{0} \!\!\!
\end{array}
\mright] \!\!\!}_{\triangleq E_{\text{c}}}  \;
	\underbrace{\!\!\!\begin{bmatrix}
		\!	\dot{\psi}_{\text{des}}(t) \!\\ \! a_{\text{ref}}(t) \!
	\end{bmatrix}\!\!\!}_{\triangleq u_{\text{ref}(t)}}\; ,
\end{multline}
where $\boldsymbol{0}$ denotes a block matrix of zeros with appropriate size. In the representation \eqref{new_rep_lin_model}, $\delta(t)$ and $a(t)$ are considered as augmented states while $\dot{\delta}(t)$ and $\dot{a}(t)$ are the control inputs. 

% \begin{comment}
% \textcolor{red}{The model is discretized using exact discretization:
% \begin{equation*}
% 	A = e ^{A_{\text{c}}\cdot t_s},
% 	\label{A_dis}
% \end{equation*}
% \begin{multline*}
% 	B_{\text{c}} = \begin{bmatrix} B_{\text{MPC-c}} & B_{\text{ref-c}}
% 	\end{bmatrix} \longrightarrow 	B = \int_{\tau = 0}^{t_s} e^{A_{\text{c}} \cdot \tau} d\tau \cdot B_{\text{c}}
% 	\label{B_dis}	\\ 
% 	= \begin{bmatrix}
% 		B_{\text{MPC}} & B_{\text{ref}},
% 	\end{bmatrix}	
% \end{multline*}
% where $t_s$ is the sampling time, $A_{\text{c}}$, $B_{\text{MPC-c}}$ and $B_{\text{ref-c}}$ are from \cref{new_rep_lin_model}. The discretized system matrices are $A$ and $B$. The discretized model can be represented as:}
% \end{comment}
Model \eqref{new_rep_lin_model} is discretized by the exact discretization method \cite{DeCarlo1989}. The discrete time linear vehicle model for a constant longitudinal vehicle velocity is used as the predictor in the MPC problem \eqref{MPC}, as follows:
\begin{equation}\label{linear_model_path_following}
	x_{i+1|k} = A x_{i|k} + B u_{i|k} + E u^{\text{ref}}_{i|k}.
\end{equation}
The state constraint \eqref{state_constraint_in_MPC}, is a polytopic set of the form $ \mathcal{X}_{i|k} = \{x_{i|k} \in \mathbb{R}^n | G^{\rm x} x_{i|k} \leq h^{\rm x}_{i|k}  \}$, where
\begin{equation}
 	G^{\rm x}\!\!= \!\begin{bmatrix}
		l_{1}  &  0  &  0 &  0 &  0 &  0 &  0 &  0\\  0  & 0  & 0  & 0  & 0  & 0  & 1  & 0 \\ 0 &  0 &  0  & 0  & 0 &  0  & -1  & 0 \\  0 &  0  & 0 &  0  & 0  & 0  & 0  & 1 \\  0&   0 &  0 &  0 &  0 &  0  & 0 &  -1
	\end{bmatrix}\!\!\!, 
		h^{\rm x}_{i|k} \!\!=\!\! \begin{bmatrix}
		\bar{h}^{\rm x_y}_{i|k} \\ \delta_{\text{max}} \\ \delta_{\text{min}} \\ a_{\text{max}} \\ a_{\text{min}}
	\end{bmatrix},
	\label{Ge}
%	\smallskip
\end{equation}
% \begin{equation}
% 	h^{\rm x}_{i|k} = \begin{bmatrix}
% 		\bar{h}^{\rm x_y}_{i|k} & \delta_{\text{max}} & \delta_{\text{min}} & a_{\text{max}} & a_{\text{min}}
% 	\end{bmatrix}^\top,
% 	\label{he}
% \end{equation}
depending on the side where the vehicle should overtake the obstacle, $l_1$ in \eqref{Ge} is chosen as $+1$ or $-1$. An obstacle avoidance is enforced in the optimization problem \eqref{MPC} by a nonzero value for $\bar{h}^{\rm x_y}_{i|k}$ in \eqref{Ge} along a prediction horizon. Consider $w^{\text{obs}}$ and $w^{\text{veh}}$ as width of the obstacle and width of the vehicle, respectively. For the steps over the prediction horizon, the existence of an obstacle is indicated by a non-zero value for $\bar{h}^{\rm x_y}_{i|k}$ as:
\begin{equation}\label{h_k}
   \bar{h}^{\rm x_y}_{i|k} = -\frac{w^{\text{obs}}}{2} - \frac{w^{\text{veh}}}{2}; 
\end{equation}
otherwise, $\bar{h}^{\rm x_y}_{i|k}$ is equal to zero.

% \begin{comment}
%  Consider ${veh}_x$, ${veh}_v$, ${veh}_{\text{width}}$, ${obs}_{x}$, ${obs}_{\text{width}}$, and ${obs}_{\text{length}}$ as the vehicle's current placement, longitudinal speed of the vehicle, the vehicle's width, placement of the obstacle, the obstacle's width, and length of the obstacle. The sampling time is denoted with $t_s$.
% Then, the moment that the vehicle will crash the obstacle $t_{\text{crash}}$ and the moment that the vehicle will pass the obstacle $t_{\text{pass}}$ are calculated as $t_{\text{crash}} = \frac{{veh}_x - \text{obs}_{x}}{{veh}_v}$, and $t_{\text{pass}} = \frac{{veh}_x - {obs}_{x} + {obs}_{\text{length}} }{{veh}_v}$. However, $t_{\text{crash}}$ and $t_{\text{pass}}$ cannot still be used in MPC. The steps that the crash and passing between the vehicle and the obstacle take place in a prediction horizon are calculated as ${step}_{\text{crash}} = \frac{t_{\text{crash}}}{t_s}$ and ${step}_{\text{pass}} = \frac{t_{\text{pass}}}{t_s}$, respectively. In this way, for a horizon of N steps, for the steps for the steps between ${step}_{\text{crash}}$ and ${step}_{\text{pass}}$, $\bar{h}^{\rm x_y}_{i|k}$ is calculated as $\bar{h}^{\rm x_y}_{i|k} = -\frac{{obs}_{\text{width}}}{2} - \frac{{veh}_{\text{width}}}{2}$; otherwise, it is equal to zero.
% \end{comment}

The input constraint \eqref{input_contsraint_in_MPC}, is a polytopic set of the form $\mathcal{U} = \{  u_{i|k} \in \mathbb{R}^m | G^{\rm u}  u_{i|k} \leq h^{\rm u} \}$ where 
\begin{align*}
      \! \!G^{\rm u} \!=\!\!  \begin{bmatrix}
		1 \!&\! -1 \!&\! 0 \!&\! 0 \\ 0 \!&\! 0 \!&\! 1 \!&\! -1\end{bmatrix}^\top  \! \! \!,   
		h^{\rm u} \!=\! \!\begin{bmatrix}
		\Delta \delta_{\text{max}} \!\!&\!\! \Delta \delta_{\text{min}} \!\!&\!\! \Delta a_{\text{max}} \!\!&\!\! \Delta a_{\text{min}}
	\end{bmatrix}^\top. 
\end{align*}
\subsection{Plant model mismatch}
The concept of a control supervisor carries two main difficulties. First, the detection lead time $t_d$ must generally be greater than a given, non-zero reaction time $t_r$. Over this time span, the control supervisor generally does not know about the inputs of the operating controller. Hence it needs to make assumptions that may not always be correct and might lead to detection errors of type 1 and type 2. This issue is not followed further in this paper, as it is assumed that the current operating control input $u_k^{\rm o}$ is precisely known to the supervisor and that the reaction time $t_r$ is less than one sampling time step.

Second, any predictions made by the control supervisor with regards to future constraint violations by the system are limited by errors in the dynamic model and uncertainty in the model parameters. In particular, using the linear model \eqref{linear_model_path_following} invariably leads to a model mismatch with the real vehicle. To account for this model error, the linear vehicle model in \eqref{linear_model_path_following} is extended as
\begin{equation}
x_{i+1|k} = A x_{i|k} + B u_{i|k} + E u^{\text{ref}}_{i|k} + w_{i|k}\;,
\end{equation}
where $w_{i|k} \in W$ is a disturbance that is used to reflect the plant-model mismatch. Here  $W\subset\mathbb{R}^8$ represents a bounded disturbance set that will be determined experimentally, as will be discussed in the following subsection.

\begin{algorithm}[t]
	\caption{The algorithm of a SOAM}
	\label{alg_safety}
	\begin{algorithmic}	[1] 
		\STATE 	$ k = 0, 1, \cdots$
		\STATE 	$ eventDetected = false$
		\FOR{$k$} 
		\IF{$eventDetected = false$} 
	%	\STATE	Path generation;
		\STATE Generate $u^{\rm o}_{k}$ ;
% 		\STATE Calculate $\hat{x}_{k+1}$ with respect to $u^{\rm o}_{k}$;
% 		\STATE $x_{0|k} = \hat{x}_{k+1}$;
		\STATE Solve the optimization problem (\ref{MPC}) with \eqref{sup_state_cons};
		\IF{ the optimization problem \eqref{MPC} is feasible}
		\STATE {Apply $u^{\rm o}_{k}$ to the vehicle};
		\STATE $u^{\text{backup}}_{k} = u^*_{0|k}$
		\ELSE 
		\STATE{Apply $u^{\text{backup}}_{k-1}$ to the vehicle};
		\STATE $eventDetected = true$;
		\ENDIF
		\ELSE
% 		\STATE $x_{0|k} = x_k;$
		\STATE Solve the optimization problem (\ref{MPC}) with \eqref{state_constraint_in_MPC};
		\STATE Apply $u^*_{0|k}$ to the vehicle;
		\ENDIF
		\ENDFOR
	\end{algorithmic}
\end{algorithm}

In this case study, the uncertainty set $W$ is used to artificially increase obstacle dimensions in the Supervisor MPC. Note that increasing the obstacle dimensions will shift the tendency of the MPC-based detection error from type 2 to type 1. In this way, even if the car is not exactly where it was predicted, based on the linear model \eqref{linear_model_path_following}, the SOAM will still be successful. Therefore, the state constraint for the Supervisor MPC is:
\begin{equation} \label{sup_state_cons}
	\tilde{\mathcal{X}}_{i|k} = \{ x_{i|k} \in \mathbb{R}^n| G^{\rm x} x_{i|k} \leq \tilde{h}^{\rm x}_{i|k}  \},
\end{equation} 
where $G^{\rm x}$ is chosen as in \eqref{Ge}, $	\tilde{h}^{\rm x}_{i|k} = \begin{bmatrix} \tilde{h}^{\rm x_y}_{i|k} & \delta_{\text{max}} & \delta_{\text{min}} & a_{\text{max}} & a_{\text{min}} \end{bmatrix}^\top$, and $\tilde{h}^{\rm x_y}_{i|k} = \bar{h}^{\rm x_y}_{i|k} + \Delta$, where $\Delta$ is a constant value added to the state constraint and $\bar{h}^{\rm x_y}_{i|k}$ as in \eqref{h_k}. An algorithm for an appropriate choice of $\Delta$ based on $W$ will be given in the next subsection. The algorithm is shown in \cref{alg_safety}.

\subsection{Determining the disturbance set}

In this section, an offline method for a suitable choice of $\Delta$ is going to be discussed. For every sampling time step $i$, consider $\bar{w}_i \in \mathbb{R}^8$ as the difference between the car's predicted states, calculated by the linear model \eqref{linear_model_path_following},  and the car's actual states, calculated by the nonlinear vehicle model \eqref{nonlinear_model}. For the implementation, it is important to recall that the states of the linear model \eqref{linear_model_path_following} are defined as the deviation between the vehicle's states and their reference values.

A bound on $\bar{w}_i$ is obtained by using a data-driven method. At first, all the states and the inputs are gridded into their possible values on the corresponding constraint set. Each of these grid points is considered as a possible initial condition. For each grid point, the linear and nonlinear models are simulated for one sampling time. Then, the difference between the vehicle’s states using the two models is calculated: $\bar{w}_i$. Since there are eight states and two control inputs that affect $\bar{w}_i$, ignoring the parameters that do not influence $\bar{w}_i$ reduces the calculation time significantly.

%\begin{algorithm}
%	\caption{Collection of $g$'s}
%	\label{g_collec}
%	\begin{algorithmic}	[1] 
%		\STATE 	Initialize $x,y, \psi,  V_x, a$
%		\STATE $\delta_{\text{max}} = $ Maximum steering angle
%        \STATE $\dot{\psi}_{\text{max}} = $ Maximum yaw angle rate
%		\STATE $V_{y,\text{max}} = $ Maximum lateral speed
%		\STATE $i = 1;$
%		\STATE $n = (2\delta_{\text{max}}+1)\times(2\psi_{\text{max}}+1)\times(V_{y,\text{max}}+1)$;
%		\STATE $M = \text{zeros}(6,n);$
%		\FOR { $   -\delta_{\text{max}}\leq \delta \leq \delta_{\text{max}} $ }
%				\FOR { $ -\dot{\psi}_{\text{max}} \leq \dot{\psi} \leq \dot{\psi}_{\text{max}} $ }
%						\FOR { $ V_{y,\text{max}} \leq V_y \leq V_{y,\text{max}} $ }	
%						%\STATE $\delta = j;$ 
%						%\STATE $\dot{\psi} = \frac{l}{t_s};$	
%						%\STATE $V_y = h;$
%			           	\STATE Calculating $X_{\text{nonlin}}$;	
%			           	\STATE Calculating $X_{\text{lin}}$;
%			           	\STATE $g = X_{\text{nonlin}} - X_{\text{lin}};$ 
%			           	\STATE $M(:,i) = g;$
%			           	\STATE $i = i +1$;	 
%			        	\ENDFOR	 
%	        	\ENDFOR	 
%		\ENDFOR
%	\end{algorithmic}
%\end{algorithm}   

After collecting the possible mismatch points, the convex hull of these points is calculated as $W = \text{conv}(\bar{w}_1,\bar{w}_2 , \cdots , \bar{w}_n)$, where $n \in \mathbb{N}$ is the number of possible grid points. However, to estimate $\Delta$, only the mismatch related to the vehicle's lateral position, $y$, is required. Then, the projection of $W$ on $y$ is calculated $\Delta = \text{proj}_y{(W)}$. In this way, a reasonable estimate of $ \Delta $ is obtained.
 
 %%%%%%%%%%%%%%%%%%%%%%%%%%%%%%%%%%%%%%%%%%%%%%%%%%%%%%%%%%%%%%%%%%%%%%%%%%%%%%%%
 
 \section{Simulation setup} 
 
A dual-track, 3DOF rigid vehicle body model from the Vehicle Dynamics Blockset in Matlab is used as the nonlinear model \eqref{nonlinear_model}. It provides a fairly realistic representation of the real dynamics of a vehicle \cite{MATLAB:2019}. It models longitudinal, lateral, and yaw motion of the vehicle. The parameters used in the simulation are shown in Table \eqref{para-table}.
 
In this paper, $u^{\rm o}_{k}$ is considered as a pure pursuit controller (PPC), which is a path following algorithm. The PPC can be interpreted as a human driver. It maintains the vehicle on the path by minimizing the deviations from the reference trajectory \cite{Woods2013}. The look-ahead time in the PPC setup is chosen as 0.5 seconds.

In the MPC setup the sampling time, $t_s$ is 0.1 seconds. The tuning parameters are chosen as $Q = \text{diag}(\begin{bmatrix} 1 & 1 & 200 & 1 & 1 & 1 & 1 & 1  \end{bmatrix})$, $R = \text{diag}(\begin{bmatrix} 1 & 1 \end{bmatrix})$, and $N = 10$. The constraints on steering angle and the acceleration are  $\delta_{\text{max}} = - \delta_{\text{min}} = 34^{\circ}$, $a_{\text{max}} = 2 \frac{\rm m}{\rm s^2}$ and $a_{\text{min}} = -6 \frac{\rm m}{\rm s^2}$. The constraints on the rate of change of the steering angle and the acceleration are $\Delta \delta_{\text{max}} = - \Delta \delta_{\text{min}} = 30^{\circ}$ and $\Delta a_{\text{max}} = - \Delta a_{\text{min}} = 30 \frac{\rm m}{\rm s^3}$.

%$e_{\text{obstacle}}$ is a parameter that is calculated to show when overtaking the obstacle how much should the vehicle deviate from the path. $e_{\text{obstacle}}$ is calculated based on the width of the obstacle and the vehicle. Depending on the distance between the vehicle and the obstacle, $e_{\text{obstacle}}$ might be zero or might have the nonzero value in different steps of an iteration in an MPC. Of course, if the obstacle is too far from the vehicle or the vehicle has already overtaken the obstacle, $e_{\text{obstacle}}$ is zero.
%%%%%%%%%%%%%%%%%%%%%%%%%%%%%%%%%%%%%%%%%%%%%%%%%%%%%%%%%%%%%%%%%%%%%%%%%%%%%%%%
\section{Results and Discussions}
\begin{figure}
	\includegraphics[width=0.49\textwidth]{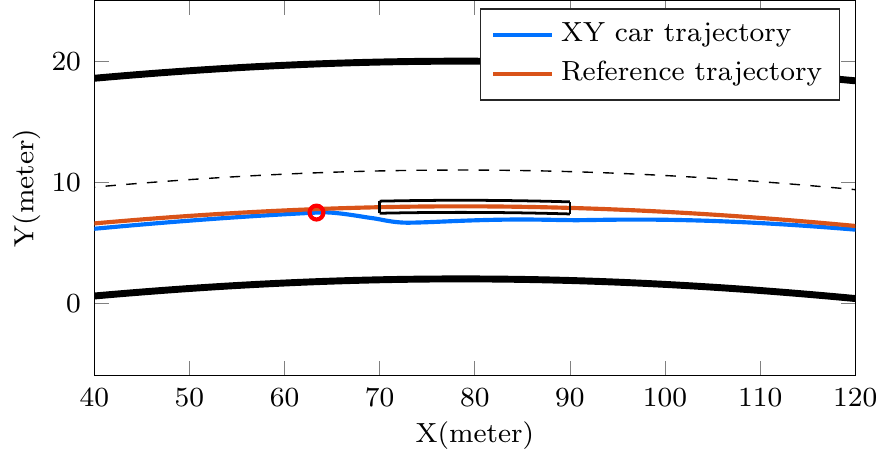}
	\caption{ A SOAM by the Backup MPC because the PPC does not see the obstacle at all.}
	\label{res_1}
\end{figure}

\begin{figure}
	\includegraphics[width=0.49\textwidth]{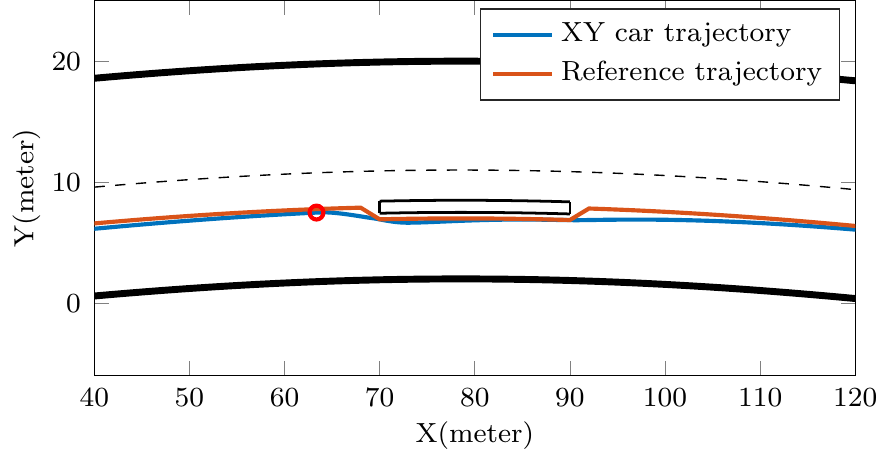}
	\caption{A SOAM by the Backup MPC because the PPC sees the obstacle too late and tries to perform an undoable maneuver.}
	\label{res_2}
\end{figure}

\begin{figure}
	\includegraphics[width=0.49\textwidth]{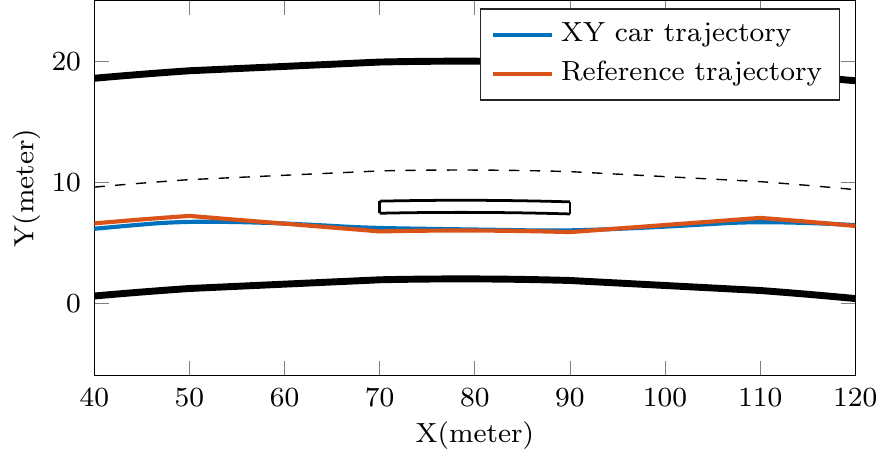}
	\caption{The Supervisor MPC does not interrupt the PPC because it is overtaking the obstacle safely.}
	\label{res_3}
\end{figure}

In the \cref{res_1,res_2,res_3}, the red line is the reference trajectory that the PPC is going to follow. The blue line is what the vehicle is doing, and the red point is where a Backup MPC starts to control the system instead of the PPC because a safety event is predicted. 

In \cref{res_1}, the PPC does not see the obstacle because of its reference trajectory. So the PPC is leading the vehicle directly into the obstacle. Nevertheless, the Backup MPC is taking control of the system at a proper time, and the vehicle is overtaking the obstacle safely.

In \cref{res_2}, the reference trajectory for the PPC includes obstacle avoidance. However, overtaking the obstacle starts so late that the SOAM is not doable due to the system's physical constraints. So again, the Backup MPC takes control of the system at a proper time and saves the vehicle from a crash.
 
In \cref{res_3}, the reference trajectory of the PPC is a smooth path. Overtaking the obstacle starts early enough. Therefore, it generates a proper control sequence for a SOAM. As shown in \cref{res_3}, the Supervisor MPC does not interrupt the PPC. Using the control input generated by the PPC, the vehicle is overtaking the obstacle safely. 

As the simulation results in \cref{res_1,res_2,res_3} confirm, the proposed SCA is working as intended for the purpose of a SOAM by this vehicle. The Supervisor MPC is generating a detection event, and when a safety event is predicted, the Backup MPC controls the vehicle for a SOAM; otherwise, the Supervisor MPC does not intervene, and the PPC controls the system for a SOAM.

% In the following section, the efficiency of choose of $\Delta$ is going to be checked because $\Delta$ is a parameter that affects the conservativeness of Supervisor MPC.
For checking the effect of $ \Delta$ on the conservatism of the SCA, introduced in Section \RomanNumeralCaps{3} part C, the result of 64 different scenarios that the vehicle is controlled by the SCA are collected. In these scenarios, it is assumed that the PPC is guiding the vehicle into the obstacle. In each scenario, two parameters are recorded. The position of the vehicle at the time when MPC switches from the Supervisor MPC to the Backup MPC, $d^{\rm MPC}$ and the position of the vehicle in the last moment before the Supervisor MPC becomes infeasible, $d^{\rm opt}$. Regarding the conservatism of the controller in these scenarios, the difference $d^{\rm MPC} - d^{\rm opt}$ is a good indicator for the quality of the chosen $\Delta$. Obviously, if $\Delta$ is chosen too large, the supervisor is more conservative, and as a result, $d^{\rm MPC} - d^{\rm opt}$ is going to be larger than otherwise. On the other hand, if $\Delta$ is too small, the vehicle might not have enough space for a safe maneuver. 

In the studied scenarios, the following reference trajectories are used during the simulations: $5 \sin{(0.008t)}+3$, $5 \sin{(0.009t)}+3$, $5 \sin{(0.01t)}+3$, $5 \sin{(0.02t)}+3$, $5 \sin{(0.03t)}+3$, $5 \sin{(0.04t)}+3$, $5 \sin{(0.05t)}+3$ and $5 \sin{(0.06t)}+3$. For each trajectory a scenario with the vehicle speeds: $5 \frac{\text{m}}{\text{s}}$, $8 \frac{\text{m}}{\text{s}}$, $10\frac{\text{m}}{\text{s}}$, $12 \frac{\text{m}}{\text{s}}$, $14\frac{\text{m}}{\text{s}}$, $16\frac{\text{m}}{\text{s}}$, $18\frac{\text{m}}{\text{s}}$ and $20  \frac{\text{m}}{\text{s}}$ is checked. 

\begin{figure}
	%width=0.5\textwidth
	%\includegraphics[scale=0.8]{Scenario_distribution.pdf}
	\includegraphics[width=0.48\textwidth]{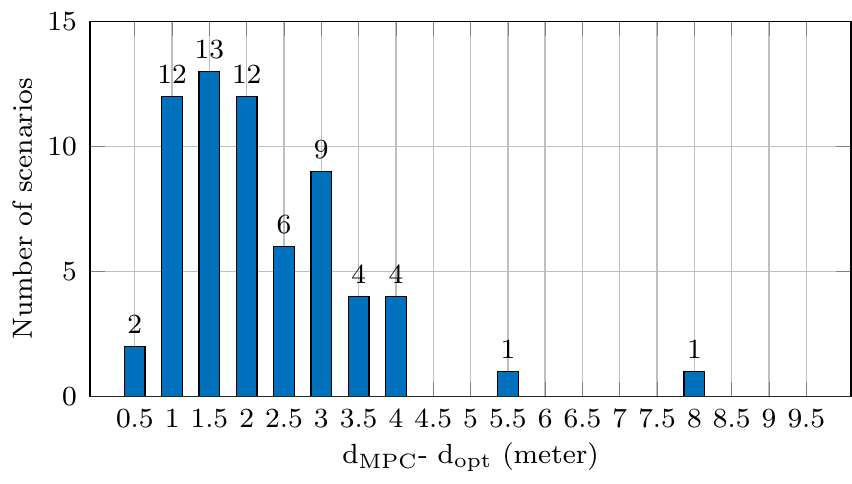}
	\caption{The distribution of $d^{\rm MPC} - d^{\rm opt}$ in 64 different scenarios}
	\label{Scenario_distribution}
\end{figure}

\begin{figure}
	%width=0.5\textwidth
	%\includegraphics[scale=0.8]{Scenario_distribution.pdf}
	\includegraphics[width=0.48\textwidth]{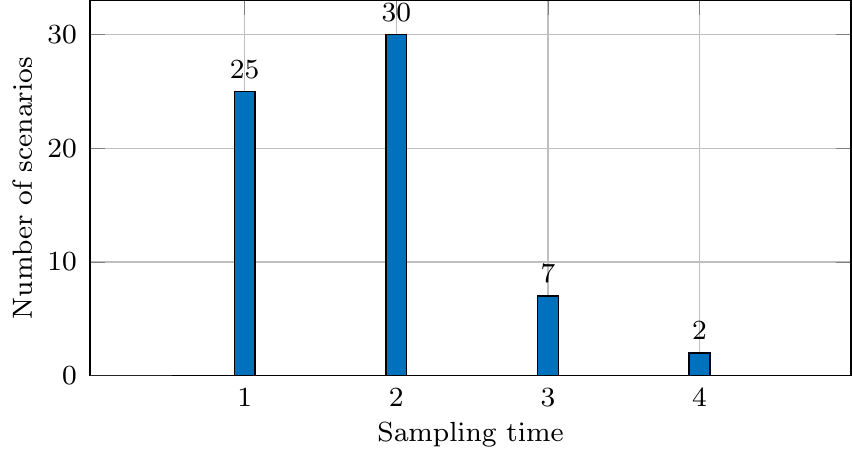}
	\caption{The distribution of the reaction time, $t_r$, based on the sampling time in 64 different scenarios}
	\label{Scenario_distribution_overtime}
\end{figure}
 The distribution of $d^{\rm MPC} - d^{\rm opt}$ for 64 scenarios is shown in \cref{Scenario_distribution}. As it is clear in \cref{Scenario_distribution}, in most of the scenarios, the Backup MPC takes over the control of the vehicle between 1 to 2 meters before the infeasibility occurs. In some cases, when the vehicle is moving at a higher speed, a safe maneuver is more difficult, so the Supervisor MPC decides to interrupt the PPC even earlier. As can be seen in \cref{Scenario_distribution}, there is no case that the Supervisor MPC is not able to predict the safety event, which is another good indicator that the choice of $\Delta$ was suitable. 
 
In \cref{Scenario_distribution_overtime}, it is shown that for the aforementioned 64 scenarios, the Supervisor MPC intervenes with the PPC between one to four sampling times earlier than the actual infeasibility occurs, while $t_d$ would ideally be only one sampling time. The greater $t_d$ is due to conservatism introduced by enlarging the obstacles' dimension by $\Delta$. Although the choice of $\Delta$ might be conservative in some scenarios, it is a suitable choice for most scenarios to ensure a successful SOAM. 

%%%%%%%%%%%%%%%%%%%%%%%%%%%%%%%%%%%%%%%%%%%%%%%%%%%%%%%%%%%%%%%%%%%%%%%%%%%%%%%%

\section{conclusions and outlook}

%textwidth : \the\textwidth \\
%The column width is: \the\columnwidth \\
%The space between columns: \the\columnsep
Ensuring the safety of a control algorithm in safety-critical systems is a real challenge. The proposed SCA in this paper provides a safety certificate for the control inputs before applying them to the system. The results show that the proposed SCA can be used for safe decision making in autonomous driving systems. The mismatch between the model and the real system leads to some problems for the proposed SCA, but they are solvable by choosing a suitable safety margin $\Delta$ when designing the SCA. \\
% The proposed method for the choice of $\Delta$ in this paper might be conservative in some scenarios, but the findings of this paper lay the ground floor for future research into this topic. 
The next step is to derive mathematical safety guarantees for the tracking problem by the proposed SCA. A further goal is to conduct experimental tests to assure the applicability of the SCA in real-world scenarios. 
%  This article has addressed safety for one of the most important safety-critical systems, safe obstacle avoidance for ground autonomous driving vehicles. To make sure that the driver is not capable of any dangerous maneuvers using an Supervisor MPC was proposed.

% The supervisor's duty was to ensure that if the driver was leading the vehicle to the obstacle, it notices early enough to take over the vehicle. To fulfill this aim, the detection of a safety event was introduced. The detection of a safety event was based on the result of the optimization problem in the supervisor. 

% Next, by using a data-driven method, the mismatch between the states of the vehicle and the linear model was collected. This data was used to establish a robust supervision method. The robust supervision was based on considering two working modes for the MPC. Using robust supervision made sure that the infeasibility in MPC happens just because of the presence of an obstacle and not because of the model error.

% In the end, some tests were carried out to make sure that the proposed methods are working effectively. In three simulations, it was shown that the supervisor interrupts the driver only when it is needed. Then 64 scenarios were tested to see when does the supervisor takes over the vehicle. The results showed that it is the most probable that it happens between 1 to 2 meters before the MPC's infeasibility, which is a good result. 

\printbibliography

\addtolength{\textheight}{-12cm}   % This command serves to balance the column lengths

%%%%%%%%%%%%%%%%%%%%%%%%%%%%%%%%%%%%%%%%%%%%%%%%%%%%%%%%%%%%%%%%%%%%%%%%%%%%%%%%

%%%%%%%%%%%%%%%%%%%%%%%%%%%%%%%%%%%%%%%%%%%%%%%%%%%%%%%%%%%%%%%%%%%%%%%%%%%%%%%%

%%%%%%%%%%%%%%%%%%%%%%%%%%%%%%%%%%%%%%%%%%%%%%%%%%%%%%%%%%%%%%%%%%%%%%%%%%%%%%%%

\end{document}